\documentclass[12pt]{article}
\usepackage{epsfig}
\newcommand{\half}{\ensuremath{\frac{1}{2}}}
\newcommand{\thalf}{\ensuremath{\frac{3}{2}}}
\newcommand{\fhalf}{\ensuremath{\frac{5}{2}}}
\newcommand{\ihalf}{\ensuremath{\frac{i}{2}}}

\newcommand{\es}{\ensuremath{(e^{2}\!-\!1)^{\frac{1}{2}}}}
\newcommand{\modx}{\ensuremath{|\mbox{\boldmath$x$}|}}
\newcommand{\modxz}{\ensuremath{|\mbox{\boldmath$x$}^{0}|}}
\newcommand{\xv}{\ensuremath{\mbox{\boldmath$x$}}}
\newcommand{\yv}{\ensuremath{\mbox{\boldmath$y$}}}
\newcommand{\Xv}{\ensuremath{\mbox{\boldmath$X$}}}
\newcommand{\xvzero}{\ensuremath{\mbox{\boldmath$x$}^{0}}}
\newcommand{\xvone}{\ensuremath{\mbox{\boldmath$x$}^{1}}}
\newcommand{\xh}{\ensuremath{\mbox{\boldmath$\hat{x}$}}}
\newcommand{\xhc}{\ensuremath{\hat{x}}}
\newcommand{\xzh}{\ensuremath{\mbox{\boldmath$\hat{x}$}^{0}}}
\newcommand{\av}{\ensuremath{\mbox{\boldmath$a$}}}
\newcommand{\ev}{\ensuremath{\mbox{\boldmath$e$}}}
\newcommand{\eone}{\ensuremath{\mbox{\boldmath$e$}_{1}}}
\newcommand{\etwo}{\ensuremath{\mbox{\boldmath$e$}_{2}}}
\newcommand{\ethree}{\ensuremath{\mbox{\boldmath$e$}_{3}}}
\newcommand{\pv}{\ensuremath{\mbox{\boldmath$p$}}}
\newcommand{\pvzero}{\ensuremath{\mbox{\boldmath$p$}^{0}}}
\newcommand{\Pv}{\ensuremath{\mbox{\boldmath$P$}}}
\newcommand{\piv}{\ensuremath{\mbox{\boldmath$\pi$}}}
\newcommand{\Ov}{\ensuremath{\mbox{\boldmath$O$}}}
\newcommand{\zerov}{\ensuremath{\mbox{\boldmath$0$}}}
\newcommand{\xd}{\ensuremath{\mbox{\boldmath$\dot{x}$}}}
\newcommand{\xzd}{\ensuremath{\mbox{\boldmath$\dot{x}$}^{0}}}
\newcommand{\xdd}{\ensuremath{\mbox{\boldmath$\ddot{x}$}}}
\newcommand{\xzdd}{\ensuremath{\mbox{\boldmath$\ddot{x}$}^{0}}}
\newcommand{\xodd}{\ensuremath{\mbox{\boldmath$\ddot{x}$}^{1}}}
\newcommand{\pd}{\ensuremath{\mbox{\boldmath$\dot{p}$}}}
\newcommand{\U}{\ensuremath{\mbox{\boldmath$U$}}}
\newcommand{\Qt}{\ensuremath{\tilde{Q}}}
\newcommand{\Pt}{\ensuremath{\tilde{P}}}
\newcommand{\Ot}{\ensuremath{\tilde{O}}}
\newcommand{\Lv}{\ensuremath{\mbox{\boldmath$L$}}}
\newcommand{\Vv}{\ensuremath{\mbox{\boldmath$V$}_{RL}}}
\newcommand{\psip}{\ensuremath{\mbox{\boldmath${\psi}_{+}$}}}
\newcommand{\psim}{\ensuremath{\mbox{\boldmath${\psi}_{-}$}}}
\newcommand{\psipd}{\ensuremath{\mbox{\boldmath${\dot{\psi}}_{+}$}}}
\newcommand{\psimd}{\ensuremath{\mbox{\boldmath${\dot{\psi}}_{-}$}}}
\newcommand{\psipi}{\ensuremath{\mbox{\boldmath${\psi}_{+}^{I}$}}}
\newcommand{\psimi}{\ensuremath{\mbox{\boldmath${\psi}_{-}^{I}$}}}
\newcommand{\psipone}{\ensuremath{\mbox{\boldmath${\psi}_{+}^{1}$}}}
\newcommand{\psimone}{\ensuremath{\mbox{\boldmath${\psi}_{-}^{1}$}}}
\newcommand{\psiptwo}{\ensuremath{\mbox{\boldmath${\psi}_{+}^{2}$}}}
\newcommand{\psimtwo}{\ensuremath{\mbox{\boldmath${\psi}_{-}^{2}$}}}
\newcommand{\psipt}{\ensuremath{\psi_{+3}}}
\newcommand{\psimt}{\ensuremath{\psi_{-3}}}
\newcommand{\psipdt}{\ensuremath{\dot{\psi}_{+3}}}
\newcommand{\psimdt}{\ensuremath{\dot{\psi}_{-3}}}
\newcommand{\di}{\ensuremath{{\partial}_{i}}}
\newcommand{\cd}{\ensuremath{\!\cdot\!}}

\begin{document}
\begin{titlepage}
\vspace{2cm}
\title{
\vskip -50pt
\begin{flushright}
\begin{normalsize}
 \ DAMTP--2002--56,
{\tt hep-th/0205232}\\ 
\end{normalsize}
\end{flushright} 
\vskip 40pt
\begin{bfseries}
The classical supersymmetric Coulomb problem
\end{bfseries}
}
\vspace{2cm}
\author{\\ R. Heumann \footnote{email: R. Heumann@damtp.cam.ac.uk}
\bigskip\\
\textit{Department of Applied Mathematics and Theoretical Physics}\\ 
\textit{University of Cambridge}\\
\textit{Wilberforce Road, Cambridge CB3 0WA, England}\\
}
\date{\large June 15, 2002}
\maketitle
\thispagestyle{empty}
\begin{abstract}
\noindent
After setting up a general model for supersymmetric classical
mechanics in more than one dimension we describe systems with centrally
symmetric potentials and their Poisson algebra. We then apply
this information to the investigation and solution of the
supersymmetric Coulomb problem, specified by an $\frac{1}{\modx}$
repulsive bosonic potential. 
\end{abstract}
\hspace{0.8cm}

\vspace{8cm}
\begin{tt}
25 pages, 2 figures
\end{tt}
\end{titlepage}
\section{Introduction}\label{introduction}

The introduction of supersymmetry into classical mechanics has been
motivated in part by the study of supersymmetric \textit{quantum}
mechanics, e.g. in the influential paper by Witten
\cite{Witten}, not the least with the aim to find a classical 
``background'' for these models. 

That this can be achieved has been long known, namely  that classical 
supersymmetric mechanics can be consistently constructed by replacing
all dynamical quantities of the theory in question by even and odd
elements of a Grassmann algebra $\mathcal{B}$ according to their
respective bosonic or fermionic nature. This Grassmann-valued
mechanics has been studied first by Berezin and Marinov \cite{Berezin}
and Casalbuoni \cite{Casalbuoni}. A more recent account devoted to
one-dimensional systems was given by Junker and Matthiesen in
\cite{Junker}. For a general review of Grassmannian geometry the
reader may turn to the book by de Witt \cite{deWitt}.

An interesting but somewhat different approach has been taken in a
series of papers by Gozzi et. al. \cite{Gozzi}. Using path integral
methods in classical mechanics he and coauthors construct an $N=2$
supersymmetric theory for any given Hamiltonian system, where fermionic
fields correspond to sections of various cotangent and tangent bundles of
the configuration manifold. It should be noted, however, that the
resulting model differs from the one analyzed in this paper,
so that solutions or symmetries cannot be directly compared.

While there is thus a considerable amount of literature on supersymmetric
mechanics and interest in the topic has grown recently, there is
a surprisingly low number of specific models that have been
investigated and solved. Furthermore, attention is usually restricted
to the case of one spatial dimension. An interesting exception on
two-dimensional Liouville systems is given in \cite{Izquierdo}.

We try in this paper to address both of these shortcomings, first by
establishing some general properties of multidimensional
supersymmetric mechanics, especially for centrally symmetric
potentials and then by analyzing and solving a particular classical
model, the Coulomb problem.

The $1/\modx$-potential is one of the oldest and most prominent
potentials that has been studied in physics. This is not only since it
properly describes both the gravitational and the electrical
interaction of massive or charged point particles, respectively but
also because it has some very attractive features. One example is that
by a theorem of Bertrand \cite{Bertrand} the attractive version of the
potential is the only centrally symmetric potential next to the
harmonic oscillator such that every admissable trajectory is
closed. Furthermore one can derive an explicit time-dependent solution
for every given set of initial data. This is a very useful property
because it implies, if perhaps unexpectedly at first sight, that the
supersymmetric version of the theory can be explicitly solved, too.

This paper is based on the experience gained from an earlier investigation
into one-dimensional supersymmetric problems. Both in \cite{HeumannManton} 
and \cite{Manton}, where a different concept of complex conjugation for 
Grassmann variables was used, one of the ideas of tackling the supersymmetric
problem was the assumption of a \textit{finitely} generated Grassmann
algebra $\mathcal{B}$. This led to a layer-by-layer structure of the
theory which could be used to solve one-dimensional models by
subsequently solving higher layers based on the solutions of all lower
ones. We shall take a similar approach in this paper.

The starting point for our investigation will be the statement of the
Lagrangian in section \ref{coulomb}. It is obtained ultimately from
the familiar $N=2$ supersymmetric $(1+1)$-dimensional field theory
with Yukawa interaction via dimensional reduction which leads to
the description of one-dimensional supersymmetric point particle
mechanics analyzed in \cite{HeumannManton}. The resulting Lagrangian
can then be generalized to more than one dimension. Notice that
supersymmetry requires the bosonic potential to be non-negative. This
means that our attention has to be restricted to the case of repulsive
interaction.

We go on to derive the complete set of $3n$ equations of motion and
show that as in the one-dimensional model there are
two independent supersymmetries as well as a purely fermionic
invariance transformation which can be seen as an internal
``rotation'' of the fermionic variables. We proceed by giving one
explicit solution to the fermionic equations of motion utilizing the
supercharges -- except in the one-dimensional case this can not be the
complete solution, though.

We then restrict our attention to a special class of models, namely
those with a centrally symmetric scalar potential. After restating
the equations of motion we derive in section \ref{centralpotentials}
the explicit form of the supersymmetric angular momentum and show that
it is a supersymmetric invariant.

The symmetries of our model can be understood best from its Poisson
algebra which we therefore describe completely in section \ref{poisson}.

In section \ref{Coulombproblem} we write down the complete set of
equations governing the supersymmetric Coulomb problem. Even in three
dimensions these are still nine coupled first and second order
differential equations and it seems by no means obvious that there
will be an explicit analytical solution for every possible set of
initial data.  

We use section \ref{solutions} to show that such an explicit solution
exists. This is achieved by utilizing a method originally employed in 
\cite{Manton} and then applied further in \cite{HeumannManton}. The
key idea is to decompose all dynamical quantities (and therefore every
equation) into terms containing the same number of Grassmann
generators. The model then naturally obtains a layer structure which
can be analyzed from the bottom layer -- the classical problem --
upwards. In particular, solutions to higher layers may be found by
using all those already derived for the lower layers. We will carry
out this programme here up to second order in the Grassmann algebra --
since our results indicate a close resemblance of the multi-dimensional 
to the one-dimensional case we can suspect that higher order solutions
exhibit many of the characteristics described already for the
one-dimensional case. We will give a short summary of these in our conclusions.

As mentioned we begin our analysis in section \ref{sectionbottom} with 
the bottom layer of the system which is identical to the classical,
i.e. non-supersymmetrized problem. Although the solutions to the
problem are well-known we explain briefly how explicit time-dependent
expressions can be found -- we do this because the textbook treatments
of the problem usually only derive the possible orbits but do not
specify the actual functions of time that describe it.

The fermionic equations of motion comprise the next layer and we study
their solutions in detail in subsection \ref{fermionic-equations}. As
in the one-dimensional case we find that fermionic motion is not only
compact but restricted to a spherical space. Furthermore, while there
is a plane of motion for the bosonic variables (of lowest order) the
same is not true for the fermionic variables or, in other words,
fermionic motion is non-trivial in all three spatial directions.

Finally, we solve the top layer of our system, consisting of the
bosonic quantities of second order in the generators. Though the
particular solutions found in the one-dimensional case cannot be
applied directly to the Coulomb problem we can carry over the most
important aspect to the three-dimensional case, namely that solutions
to the homogeneous equations of motion of the top layer system are
\textit{variations} of the bottom layer system with the free
parameters of that motion. In particular, we show that 
every top layer solution to the homogeneous equations of the Coulomb
problem can be explained as a linear combination of the variations
with initial time, energy, eccentricity, orientation of angular
momentum and orientation of the hyperbola in the plane of motion.
We conclude our analysis by a full statement of the explicit solutions
of the inhomgeneous equations including the boson-fermion-interaction term.

As is well-known, the classical Coulomb problem exhibits a hidden
$O(3,1)$-symmetry as a consequence of which there is an extra
vector-valued conserved quantity, the Runge-Lenz-vector. It comes as
an unexpected surprise that this symmetry is broken in the supersymmetric 
version of the problem. We demonstrate this, i.e. that there can
be no supersymmetric version of the Runge-Lenz vector, in the appendix.

\section{Multidimensional supersymmetric mechanics}\label{coulomb}
We assume that supersymmetric mechanical models can be described by
the following Lagrangian:
\begin{equation}
\mathcal{L}=\half\xd\cd\xd-\half\U(\xv)\cd\U(\xv)
+\frac{i}{2}\psip\cd \psipd
+\frac{i}{2}\psim\cd \psimd
+i\psip\nabla\U(x)\psim,
\label{Lagrangian}
\end{equation}
which is a generalization of the familiar supersymmetric Lagrangian in
one-dimensional mechanics \cite{HeumannManton}, ultimately derived
from $(1\!+1\!)$-dimensional field theory. The dynamical quantities $\xv$,
$\psip$ and $\psim$ are $n$-component vectors, e.g. $\xv=(x_{1},\dots,x_{n})$, 
$\xv$ is Grassmann-even and $\psip$ and $\psim$ are Grassmann-odd.

In addition, $\U$ is taken to be a vector-valued $n$-component Grassmann-even
function of $\xv$, so that $\nabla\U$ is a $n\times n$-tensor with
elements $\di U_{j}(x_{k})$. We will usually take this tensor to be
symmetric which will always be the case if $\U$ is derived from a
scalar potential term as $\U(\xv)=\nabla W(\xv)$.

$\xv\cd\mbox{\boldmath$y$}$ denotes the standard Euclidean inner
product, extended if necessary in the obvious way:
$\xv\mbox{\boldmath$M$}\mbox{\boldmath$y$}$,
\mbox{\boldmath$M$} being an $n\times n$-tensor, stands e.g. for the 
component expression $x^{i}M_{ij}y^{j}$.

\bigskip
The equations of motion can now be read off from the Lagrangian:
\begin{eqnarray}
\xdd\:\:&=&-\nabla\U(\xv)\U(\xv)+
i\nabla\left(\psip\nabla\U(\xv)\psim\right)\label{boson-equation}\\
\psipd&=&-\nabla\U(\xv)\psim\label{fermion-equation 1}\\
\psimd&=&\quad\nabla\U(\xv)\psip.\label{fermion-equation 2}
\label{eqns of motion}
\end{eqnarray}
Every equation is vector-valued and has $n$ components.

\bigskip
Invariance with respect to time-translation means that there is a
conserved Hamiltonian which can be calculated to be
\begin{equation}
H=\half\xd\cd\xd+\half\U(\xv)\cd\U (\xv)
-i\psip\nabla\U(\xv)\psim.
\label{Hamiltonian}
\end{equation}
In addition there are two independent supersymmetry transformations,
namely
\begin{equation}
\begin{array}{rclrclrcl}
\delta \xv &=& i \epsilon \psip,& \delta \psip&=& -\epsilon \xd,&
\delta \psim&=&-\epsilon \U(\xv),\\
\tilde{\delta} \xv &=& i \epsilon \psim,&\tilde{\delta} \psip&=& 
\:\:\:\,\epsilon \U(\xv),&\tilde{\delta} \psim&=&-\epsilon \xd,\label{susy}
\end{array}
\end{equation}
where $\epsilon$ is an infinitesimal scalar Grassmann-odd parameter. The
corresponding conserved supercharges are given by the expressions
\begin{eqnarray}
Q&=&\xd\cd\psip+\U(\xv)\cd\psim\label{Q}\\
\Qt&=&\xd\cd\psim-\U(\xv)\cd\psip.\label{Qtilde}
\end{eqnarray}
Finally, an internal transformation in the fermionic part of the
Grassmann algebra, given by
\begin{displaymath}
\delta\psip=\eta\psim,\quad\delta\psim=-\eta\psip,
\end{displaymath}
where $\eta$ denotes an infinitesimal scalar Grassmann-even
parameter, also leaves the Lagrangian invariant. The resulting charge is simply
\begin{equation}
R=i\psip\cd\psim.\label{R}
\end{equation}

One of the results from \cite{HeumannManton}, that can
be carried over from the one-dimensional case almost without
alteration, is that solutions to the fermionic equations can be found
using the supercharges. In one dimension
(\ref{Q}) and (\ref{Qtilde}) can be formally inverted to yield:
\begin{eqnarray}
\psip&=&\frac{1}{2E}(Q\xd-\Qt\U)\label{fermion1-solution}\\
\psim&=&\frac{1}{2E}(Q\U+\Qt\xd)\label{fermion2-solution},
\end{eqnarray}
where $E$ is the constant Grassmann-valued energy. This is still a
solution in the $n$-dimensional case, though special care has to
be taken if the real part $E_{0}$ is zero. 

Unlike the situation in one dimension, however, (\ref{fermion1-solution}) and
(\ref{fermion2-solution}) give only part of the fermionic solution. In
the multi-dimensional case there is still a $2(n\!-\!1)$-dimensional solution
space that needs to be determined.

\section{Centrally symmetric potentials and angular momentum}
\label{centralpotentials}
So far these formulae have been straightforward generalizations of formulae
from the one-dimensional case. However, in more than one dimension there can be
additional conserved quantities if there are extra symmetries in the
system. We will now assume that the scalar bosonic potential
\begin{displaymath}
V=\half\,\U(\xv)\cd\U(\xv)
\end{displaymath}
is central, i.e. a function of $\modx=(\xv\cd\xv)^{\half}$ only. This
can be easily achieved by demanding that the scalar potential $W(\xv)$
is a function of $\modx$, for then $\U(\xv)$ will be of the form
\begin{displaymath}
\U(\xv)=W'(\modx)\xh,
\end{displaymath}
with $\xh$ denoting the unit vector $\xv/\modx$.

The equations of motion can now be written as follows:
\begin{eqnarray}
\xdd\:\:&=&-W'W''\xh
+i\left(\frac{W''\modx-W'}{\modx^{2}}\right)
\left(\psip(\psim\cd\xh)+(\psip\cd\xh)\psim\right)
\label{boson-equation-central}\\
&&+i\left[\left(\frac{W'''\modx^{2}-3(W''\modx-W')}{\modx^{2}}\right)
(\psip\cd\xh)(\psim\cd\xh)+\left(\frac{W''\modx-W'}{\modx^{2}}\right)
(\psip\cd\psim)\right]\xh\nonumber\\
\psipd&=&-\left(\frac{W'}{\modx}\right)\psim
-\left(\frac{W''\modx-W'}{\modx}\right)(\psim\cd\xh)\xh
\label{fermion1-equation-central}\\
\psimd&=&\quad\left(\frac{W'}{\modx}\right)\psip
+\left(\frac{W''\modx-W'}{\modx}\right)(\psip\cd\xh)\xh,
\label{fermion2-equation-central}
\end{eqnarray}
where the argument of $W$ is $\modx$. Note that the purely bosonic
part of the problem is described just by the first term on the right
hand side of (\ref{boson-equation-central}).

\bigskip
For convenience we will now restrict ourselves to the
three-dimensional case relevant for the Coulomb problem, although our
discussion should generalize easily to more than three dimensions.

The particular form of the potential function $W$ means that
there is an extra symmetry in the system, given by rotating
both the bosonic and the fermionic variables by the same
amount around the same axis $\av$. Infinitesimally, we can write
\begin{displaymath}
\delta_{a}\xv=\epsilon[\av,\xv];\quad
\delta_{a}\psip=\epsilon[\av,\psip];\quad
\delta_{a}\psim=\epsilon[\av,\psim],
\end{displaymath}
where $[\cdot,\cdot]$ denotes the three-dimensional vector product,
i.e. $[\xv,\yv]^{i}=\epsilon_{ijk}x_{j}y_{k}$.

This leads to the conserved angular momentum:
\begin{equation}
\Lv=\left[\xv,\xd\right]-\ihalf\left[\psip,\psip\right]
-\ihalf\left[\psim,\psim\right].
\end{equation}
Note that the second and third term of $\Lv$ are non-zero since
$\psip$ and $\psim$ are Grassmann-odd.

It is an interesting observation that the strict condition for the
existence of angular momentum is that $W$ is central. By choosing 
$\U=f(\modx)\,\Ov\xv;\Ov\!\in\!O(3)$ we end up with a rotationally invariant 
bosonic potential $V$ but angular momentum is not conserved.

\bigskip
Finally we point out that the following two expressions are also
time-independent:
\begin{displaymath}
\begin{array}{c}
[\psip,\psip]\cd[\psim,\psim]\\
\left([\psip,\psip]\cd\psip\right)\cd\left([\psim,\psim]\cd\psim\right),
\end{array}
\end{displaymath}
which can be said to denote the product of the oriented ``areas'' and
``volumes'' of $\psip$ and $\psim$, respectively.
They do not, however, constitute independently conserved quantities as
the first one is proportional to $R^{2}$ and the second one to
$R^{3}$.

\bigskip
Apart from invariance under time translation, another important
aspect is that of invariance under supersymmetry. It can be shown
directly that angular momentum and energy are supersymmetric
invariants whereas e.g. the extra charge $R$ is not:
\begin{equation}
\begin{array}{rclrclrclrclrcl}
\delta H &=&0, & \delta \Lv &=& \zerov,&
\delta Q &=&-2\epsilon H,&\delta\Qt&=&\quad 0,&
\delta R &=&-i\epsilon\Qt\\
\tilde{\delta} H &=&0,&\tilde{\delta} \Lv&=& \zerov,&
\tilde{\delta} Q &=&\quad 0,&\tilde{\delta}\Qt&=&-2\epsilon H,&
\tilde{\delta} R &=&\:\:\:i\epsilon Q.
\end{array}
\end{equation}
Here $\delta$ and $\tilde{\delta}$ are the two supersymmetry
transformations introduced in equation (\ref{susy}) and
$\epsilon$ is a Grassmann-odd infinitesimal parameter. 
\section{Poisson-Algebra}\label{poisson}
For a better understanding of the symmetries of our mechanical system
we will now derive the Poisson-algebra of the problem. Therefore, we
must write down the definition of the Poisson brackets for
Grassmann-valued quantities. To do this we have to introduce the
canonical momenta $\pv$ and $\piv$ by
\begin{displaymath}
\pv=\frac{\partial L}{\partial\xd},
\quad\piv_{+}=\frac{\partial L}{\partial\psipd},
\quad\piv_{-}=\frac{\partial L}{\partial\psimd}.
\end{displaymath}
For the specific choice of our model one finds immediately that
\begin{equation}
\pv=\xd,\quad\piv_{+}=-\ihalf\psip,\quad\piv_{-}=-\ihalf\psim.
\label{constraint}
\end{equation}
The fact that the canonical momenta associated with the fermionic
variables are up to a constant factor identical to these variables
themselves means that $\psipd$ and $\psimd$ cannot be expressed uniquely 
as a function of configuration and momentum variables 
separately. It is therefore sensible to view (\ref{constraint}) as a constraint
and replace all occurences of $\piv_{+}$ and $\piv_{-}$ accordingly.

Hamiltons equations can then be written as
\begin{displaymath}
\xd=\frac{\partial H}{\partial\pv};
\quad\pd=-\frac{\partial H}{\partial\xv};
\quad\psipd=-i\frac{\partial H}{\partial\psip};
\quad\psimd=-i\frac{\partial H}{\partial\psim},
\end{displaymath}
so that the Poisson brackets for even quantities $y$ and odd
quantities $\theta$ have to be:
\begin{eqnarray*}
\{y_{1},y_{2}\}&=&\left(
\frac{\partial y_{1}}{\partial x_{i}}\frac{\partial y_{2}}{\partial p_{i}}-
\frac{\partial y_{2}}{\partial x_{i}}\frac{\partial y_{1}}{\partial p_{i}}+ i 
\frac{\partial y_{1}}{\partial\psip_{i}}\frac{\partial y_{2}}{\partial\psip_{i}}+ i
\frac{\partial y_{1}}{\partial \psim_{i}}\frac{\partial y_{2}}{\partial \psim_{i}}
\right)\\
\{\theta,y\}&=&\left(
\frac{\partial \theta}{\partial x_{i}}\frac{\partial y}{\partial p_{i}}-
\frac{\partial y}{\partial x_{i}}\frac{\partial \theta}{\partial p_{i}}- i 
\frac{\partial \theta}{\partial \psip_{i}}\frac{\partial y}{\partial \psip_{i}}- i
\frac{\partial \theta}{\partial \psim_{i}}\frac{\partial y}{\partial \psim_{i}}
\right)\\
\{\theta_{1},\theta_{2}\}&=&\left(
\frac{\partial \theta_{1}}{\partial x_{i}}
\frac{\partial \theta_{2}}{\partial p_{i}}+
\frac{\partial \theta_{2}}{\partial x_{i}}
\frac{\partial \theta_{1}}{\partial p_{i}}- i 
\frac{\partial \theta_{1}}{\partial \psip_{i}}
\frac{\partial \theta_{2}}{\partial \psip_{i}}- i
\frac{\partial \theta_{1}}{\partial \psim_{i}}
\frac{\partial \theta_{2}}{\partial \psim_{i}}
\right),
\end{eqnarray*}
where summation over $i$ is implied. The result is the same as 
in \cite{Casalbuoni} if the constraint (\ref{constraint}) is used to
replace the occurences of $\piv_{+}$ and $\piv_{-}$ in the formulae
given there.

We can then calculate the brackets between the dynamical variables
and find the only non-zero elements:
\begin{displaymath}
\{x_{i},p_{j}\}=\delta_{ij};\quad
i\{\psip_{i},\psip_{j}\}=\delta_{ij};\quad
i\{\psim_{i},\psim_{j}\}=\delta_{ij}.
\end{displaymath}
From these we can finally determine the following algebra relations
between all the conserved quantities:
\begin{displaymath}
\begin{array}{l}
\{H,H\}=\{H,Q\}=\{H,\Qt\}=\{H,R\}=\{H,\Lv\}=0,\\
i\{Q,Q\}=i\{\Qt,\Qt\}=2H,\:\{Q,\Qt\}=0,\:
\{Q,R\}=\Qt,\:\{\Qt,R\}=-Q,\\
\{\Lv,Q\}=\{\Lv,\Qt\}=\{\Lv,R\}=0,\:
\{L_{i},L_{j}\}=\epsilon_{ijk}L_{k},\:\{R,R\}=0.
\end{array}
\end{displaymath}

\section{The Coulomb problem}\label{Coulombproblem}
The supersymmetric Coulomb problem, characterized by an $1/\modx$
repulsive bosonic potential, can now be obtained from (\ref{Lagrangian}) 
by setting
\begin{equation}
\U(\xv)=\frac{2^{\half}}{{\modx}^{\half}}\xh,\label{Coulombpotential}
\end{equation}
or equivalently by choosing $W(\modx)=2^{\thalf}\modx^{\half}$ in 
(\ref{boson-equation-central})--(\ref{fermion2-equation-central}). It
follows that the equations of motion now read:
\begin{eqnarray}
\xdd&=&\frac{\xh}{\modx^{2}}-i\thalf\frac{2^{\half}}{\modx^{\fhalf}}\left(
\psip(\psim\cd\xh)\!+\!(\psip\cd\xh)\psim\!+\!(\psip\cd\psim)\xh
\!-\!\frac{7}{2}(\psip\cd\xh)(\psim\cd\xh)\xh\right)\label{bose-coulomb}\\
\psipd&=&-\frac{2^{\half}}{\modx^{\thalf}}
\left(\psim-\thalf(\psim\cd\xh)\xh\right)
\label{fermi1-coulomb}\\
\psimd&=&\:\:\frac{2^{\half}}{\modx^{\thalf}}
\left(\psip-\thalf(\psip\cd\xh)\xh\right)
\label{fermi2-coulomb}
\end{eqnarray}
We will proceed to solve these equations for a Grassmann algebra with
two generators in the next section.

The formulae for all the conserved quantities need not be written out
again for our special choice of $\U$ but it is worth mentioning the
explicit form of the Hamiltonian:
\begin{equation}
H=\half\xd\cd\xd+\frac{1}{\modx}-i\,\frac{2^{\half}}{\modx^{\thalf}}
\left(\psip\cd\psim-\thalf(\psip\cd\xh)(\psim\cd\xh)\right).
\label{Coulomb-Hamiltonian}
\end{equation}

In the purely bosonic case the Coulomb potential admits a further well-known
conserved quantity, the Runge-Lenz vector, given by:
\begin{displaymath}
\Vv=[\Lv,\xd]-\xh
\end{displaymath}
Unfortunately this extra conserved quantity does not seem to exist for
the supersymmetric case. We will outline a proof for this in the appendix.

\section{Explicit solutions}\label{solutions}
We will now solve equations (\ref{bose-coulomb})--(\ref{fermi2-coulomb}) 
for the case of a Grassmann algebra with two generators. In other
words, we assume that all physical quantities take their values in 
an algebra that is spanned by two Grassmann-odd elements $\xi_{1}$ and
$\xi_{2}$, satisfying the relations
\begin{displaymath}
\xi_{1}^{2}=\xi_{2}^{2}=0,\:\xi_{1}\xi_{2}=-\xi_{2}\xi_{1}. 
\end{displaymath}
Using a method similar to the one deployed in \cite{HeumannManton} we
split the bosonic variable $\xv$ into components according to the number
of generators involved:
\begin{displaymath}
\xv(t)=\xvzero(t)+\xv^{ij}(t)i\xi_{i}\xi_{j}\equiv\xvzero(t)+\xvone(t).
\end{displaymath}
Taylor-expanding the potential function $\U$ then gives us
\begin{displaymath}
\U(\xv)=\U(\xvzero)+\nabla\U(\xvzero)\xvone.
\end{displaymath}
Notice that all quantities are vector-valued. We will indicate
components, if necessary, by lower indices.

\bigskip
The fermionic variables are time-dependent multiples of $\xi_{1}$ and
$\xi_{2}$ since for the particular Grassmann algebra chosen there can be no
product of three or more generators. Therefore in products of bosonic
and fermionic quantities only bosonic terms of zeroth order in the
generators contribute. Writing
\begin{eqnarray*}
\psip(t)&=&\psipone(t)\xi_{1}+\psiptwo(t)\xi_{2}\\
\psim(t)&=&\psimone(t)\xi_{1}+\psimtwo(t)\xi_{2}
\end{eqnarray*}
both fermionic variables decompose into two terms with \textit{real-valued}
time-dependent coefficients $\psipi$ and $\psimi$, respectively, where
$I=1,2$.

The equations of motion for the coefficients are similar to
(\ref{fermi1-coulomb}) and (\ref{fermi2-coulomb}) --- the only
differences being that $\psip$ and $\psim$ have to be replaced by the
component functions $\psipi$ and $\psimi$ and $\xv$ has to be replaced by
$\xvzero$.

While the fermionic equations thus do not look much different after the
decomposition the bosonic equation (\ref{bose-coulomb}) splits in two, namely
\begin{eqnarray}
\xzdd&=&\frac{\xzh}{\modxz^{2}}\label{bottom}\\
\xodd&=&\frac{\xvone-3(\xvone\cd\xzh)\xzh}{\modxz^{3}}
+\av(\xvzero,\psip,\psim),\label{top}
\end{eqnarray}
where
\begin{displaymath}
\av(\xvzero,\psip,\psim)=
-i\thalf\frac{2^{\half}}{\modxz^{\fhalf}}\left(
\psip(\psim\cd\xzh)\!+\!(\psip\cd\xzh)\psim\!+\!
(\psip\cd\psim)\xzh\!-\!\frac{7}{2}(\psip\cd\xzh)(\psim\cd\xzh)\xzh\right)
\end{displaymath}
does not contain $\xvone$. 

Following \cite{HeumannManton} we will now systematically solve these
equations by working ``upwards'' from the bottom layer of the system
(\ref{bottom}) through the fermion equations (\ref{fermi1-coulomb})
and (\ref{fermi2-coulomb}) to the top layer (\ref{top}). 

\subsection{The bottom layer bosonic equation}\label{sectionbottom}
The lowest order equation (\ref{bottom}) is of course just the
familiar equation of the Coulomb problem. Its solutions are the
well-known hyperbolae in the plane orthogonal to the angular momentum
vector $\Lv$ (if we set apart for a moment the special case $\Lv=0$).
In order to make use of the solutions for the remaining equations it will be
necessary to write them down in an explicit time-dependent
fashion. To do this we utilize a method first devised by Moser in \cite{Moser}
for the Kepler problem with a negative total energy and then extended
to the positive energy case by Belbruno \cite{Belbruno}.

We can summarize these results adapted to our problem by saying that there
is a diffeomorphism between the constant energy surface $H=E^{0}$ in the
phase space of the Coulomb problem and the tangent bundle of the upper
(or lower) sheet of the hyperboloid $\mathcal{H}$ specified in
four-dimensional Euclidean space by 
$X_{0}^{2}-X_{1}^{2}-X_{2}^{2}-X_{3}^{2}=1$. This diffeomorphism takes
geodesics on $\mathcal{H}$ into solutions to (\ref{bottom}) requiring
only a change of the time variable.
 
More explicitly, if we parametrize a geodesic on $\mathcal{H}$ by
$X_{i}(s);\;i=0,\!\dots\! ,3$ and denote by $P_{i}(s);\,\mbox{$i=0,\dots,3$}$ 
its tangent, then
\begin{eqnarray}
\xvzero(s) &=&\Pv(s)(1+X_{0}(s))-\Xv(s) P_{0}(s)\\
\pvzero(s) &=&\Xv(s)\frac{1}{1+X_{0}(s)}
\end{eqnarray}
transforms this geodesic into a solution of the Kepler problem for the
energy $E^{0}=\half$, when we assume the following transcendental
relationship between the geodesic ``time''-parameter $s$ and physical time $t$:
\begin{equation}
t=\int_{0}^{s}|\xvzero(s')|\,ds'.
\end{equation} 
Solutions for arbitrary energy can then be obtained by the scaling
\begin{equation}
\tilde{\xv}^{0}=\frac{1}{2E^{0}}\xvzero,\quad
\tilde{\pv}^{0}=(2E^{0})^{\half}\pvzero,
\quad\tilde{t}=\frac{1}{(2E^{0})^{\thalf}}t.\label{scaling}
\end{equation}

Every geodesic on $\mathcal{H}$ can be mapped using an
appropriate $SO(3,1)$-transformation into
\begin{displaymath}
\Xv(s)=(\cosh s \cosh \beta,\sinh s,0,-\cosh s \sinh \beta).
\end{displaymath}
For the Coulomb problem this equates to choosing the plane of motion
and the orientation of the hyperbola in it. (Note that for the
specific parametrization chosen we have also fixed the origin of
time.) Specifically we get
\begin{eqnarray}
\xvzero&=&(e+\cosh s, -\es\sinh s,0)\label{boson-solution}\\
\pvzero&=&(\frac{\sinh s}{1+e\cosh s},
-\frac{\es\cosh s}{1+e\cosh s},0),
\end{eqnarray}
where $e=\cosh \beta$ represents the eccentricity of the
hyperbola. The relationship between $s$ and $t$ is given by 
\begin{equation}
t=e\sinh s+s.
\end{equation}
This equation cannot be inverted analytically. 

One can read off from (\ref{boson-solution}) that the motion is
indeed hyperbolic, taking place in the \mbox{$x_{1}$--$x_{2}$-plane} with the
focal points on the $x_{1}$-axis and the origin of time chosen such
that $t=0$ at the point of closest approach to the potential
centre. We just mention that $\Lv$ is oriented in $(-x_{3})$-direction
and (for constant energy) $|\Lv|=\es$; the scattering angle $\theta$ is
determined by $e \sin \frac{\theta}{2}=1$.   

There are two special cases worth mentioning, namely $e=1$ when the
scattering angle $\theta$ is $\pi$ and the angular momentum is zero and
$e\longrightarrow\infty$ when $\theta=0$ and the equations describe
free motion at an infinite distance to the center of the potential.

\subsection{The fermionic equations}\label{fermionic-equations}
To solve equations (\ref{fermi1-coulomb}) and (\ref{fermi2-coulomb})
for $\xv\equiv\xvzero$ we can first go back to our general solutions
(\ref{fermion1-solution}) and (\ref{fermion2-solution}). They state in
this circumstance that $\psip$ and $\psim$ are linear combinations of
$\xzd$ and $\U(\xvzero)$ with Grassmann-odd coefficients. (Note that we
can safely assume $E^{0}>0$ due to the repulsive nature of the Coulomb
potential.) 

To analyse the fermionic movement it is
therefore sensible to look at the motion of $\xzd$ and
$\U(\xvzero)$. The first interesting aspect is that if we combine
velocity and potential in a six-dimensional vector
$Z=(\xzd,\U)$ then the motion of this vector takes place
on $S^{3}$, since
\begin{displaymath}
\xzd\cd\xzd+\U\cd\U=2E^{0}=1
\end{displaymath}
for our particular solution, but $\dot{x}^{0}_{3}\!=\!U_{3}\!=\!0$ trivially.
This is in fact very similar to the one-dimensional problem where
motion takes place on $S^{1}$ instead \cite{HeumannManton}.

\bigskip
If we project the motion into the $\dot{x}^{0}_{1}\!-\!\dot{x}^{0}_{2}$-plane
then this two-dimensional vector describes a circle with
center at $(0,-e/\es)$ (see Fig. 1). The motion starts for
$t=-\infty$ at the intersection of this circle with $S^{1}$, running
initially towards the origin but then bending towards the
$\dot{x}^{0}_{2}$-axis and reaching its nearest point to the origin at
$(0,-\es/(e+1))$. Then it turns back outwards and
runs symmetrically to the other intersection point, reaching it at
$t=\infty$.

\begin{figure}[ht]
\begin{center}
\epsfig{file=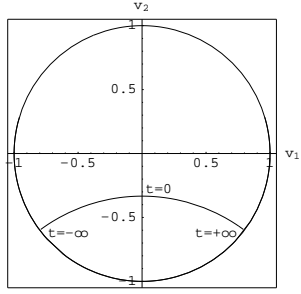,width=4cm}
\caption{{\small \it Projection of the bosonic motion into the
$\dot{x}^{0}_{1}\!-\!\dot{x}^{0}_{2}$-plane $(e=\frac{5}{4})$}} 
\label{diagram1}
\end{center}
\end{figure}

The projection into the $U_{1}\!-\!U_{2}$-plane shows a different
picture. Due to the factor $1/\modxz^{\half}$ in
(\ref{Coulombpotential}) the motion starts and ends at the origin. It
runs radially outwards along an asymptote determined by the angular
momentum but then bends towards the $U_{1}$-axis, crossing it at time
$t=0$ and returns to the origin in a symmetric way (see Fig. 2).

\begin{figure}[ht]
\begin{center}
\epsfig{file=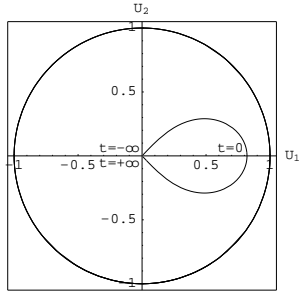,width=4cm}
\caption{{\small \it Projection of the bosonic motion into the
$U_{1}\!-\!U_{2}$-plane $(e=\frac{5}{4})$}} 
\label{diagram2}
\end{center}
\end{figure}

When we come back to solutions (\ref{fermion1-solution}) and
(\ref{fermion2-solution}) we can immediately state that
\begin{eqnarray*}
\lim_{t\rightarrow\infty}\psip=\lim_{t\rightarrow -\infty}\psip=Q\xzd\\
\lim_{t\rightarrow\infty}\psim=\lim_{t\rightarrow -\infty}\psim=\Qt\xzd,
\end{eqnarray*}
so that we know the asymptotic behaviour of these solutions.

\bigskip
However, we can visualize the fermionic motion completely if we refer
to the bosonic motion on $S^{3}$. For this it is more convenient to
use the real-valued components $\psipi$ and $\psimi$ than the
Grassmann-valued quantities. One advantage is that we can meaningfully
define the length of these vectors: For example,
\begin{displaymath}
S^{I}=\half(\psipi)^{2}+\half(\psimi)^{2}
\end{displaymath}
is a conserved quantity arising from invariance of the component equations 
of motion under change of generators. Note that because we are dealing
with real-valued quantities now, none of the two products is zero. 

To understand the close connection between the above-mentioned bosonic
motion on $S^{3}$ and fermionic motion let us ignore the third components
$\psip^{I}_{3}$ and $\psim^{I}_{3}$ for a moment, since, as we will later see,
their equations decouple. Combining the remaining fermionic terms
into the vector-valued object
\begin{displaymath}
\Psi^{I}=(\psip^{I}_{1},\psip^{I}_{2},\psim^{I}_{1},\psim^{I}_{2})
\end{displaymath}
we see that $\Psi^{I}$ lies in the plane spanned by the bosonic vector
$Z=(\xzd, \U)$ and one specific orthogonal vector $Z_{\perp}=(-\U,\xzd)$
obtained by rotating $Z$ by ninety degrees in the
$\dot{x}_{i}-U_{i}$-planes for $i\!=\!1,2$. (We are neglecting the
trivial third components here). The angle between $\Psi^{I}$ and the
two bosonic vectors is entirely specified by the two supercharges $Q$
and $\Qt$. (To be more precise, it is specified by their real-valued 
components $Q^{I}$ and $\Qt^{I}$ with respect to the two Grassmann generators. 
For simplicity, we will not indicate this difference in our notation
assuming that the precise meaning is always clear from the context.) In
other words, as the bosonic vector $Z$ moves on $S^{3}$, determined by
the bosonic equation of motion, the fermionic vector simply follows
this movement rigidly, its direction being fixed with respect to the
bosonic vector and its orthogonal complement by the two supercharges. 

\bigskip
This result is very similar to the one-dimensional case where one
finds similarly constructed bosonic and fermionic vectors corotating
on $S^{1}$. Still, there is a major difference in more than one
dimension. This lies in the fact that $Q$ and $\Qt$ do not specify the
fermionic solutions uniquely. The solution in one dimension can be
derived by inverting equations (\ref{Q}) and (\ref{Qtilde}), yet in two
or more dimensions this inversion is no longer possible. Instead these
two equations only specify the two angles between $\Psi^{I}$ and the
\textit{two} bosonic vectors $Z,Z_{\perp}$ but not more. This means that the
general $\Psi^{I}$ (again neglecting third components) moves simultaneously on 
two regular four-dimensional cones with center at the origin and their
symmetry axes $Z$ and $Z_{\perp}$, respectively. As $Z$ and $Z_{\perp}$ 
move with time $\Psi^{I}$  follows rigidly attached to the surface of both
cones and thus lies on the intersection of them. In the particular
solution that we gave above this intersection corresponds just to one
point (taken at fixed distance from the origin) -- thus specifying
this particular solution completely. However, in general the intersection 
of the two cones (and $S^{3}$) will yield a one-dimensional sphere (just 
as the intersection of two three-dimensional cones and $S^{2}$ is equivalent
to a zero-dimensional sphere, i.e. two discrete points). All points on this
sphere have the same combination of supercharges $Q$ and $\Qt$, which means
that there exists a one-dimensional space of degrees of freedom for
the fermionic motion that is not determined by the supercharges.

\bigskip
In fact, we will now derive a second solution to (\ref{fermi1-coulomb}) 
and (\ref{fermi2-coulomb}) that fixes the motion of $\Psi^{I}$ on
this one-dimensional space. Since the particular solution involving
the supercharges described above lies entirely in the
$Z$-$Z_{\perp}$-plane we will now look for a solution in the plane
orthogonal to it. (Remember, that the space we are considering is
four-dimensional.) This plane can be spanned by the two vectors:
\begin{eqnarray*}
Y&=&(\dot{x}^{0}_{2},-\dot{x}^{0}_{1},-U_{2},U_{1})\\
Y_{\perp}&=&(U_{2},-U_{1},\dot{x}^{0}_{2},-\dot{x}^{0}_{1}).
\end{eqnarray*}
A straightforward ansatz for the second solution is $\Psi=PY+\Pt
Y_{\perp}$, where $P$ and $\Pt$ are Grassmann-odd constants.
Yet, this ansatz turns out to be too simple and does not yield a
solution. Instead we have to allow for time-dependent coefficients of
$Y$ and $Y_{\perp}$. In other words we make the assumption that
\begin{equation}
\Psi=\lambda_{1}(t)Y+\lambda_{2}(t)Y_{\perp},\label{fermion2-solution-general}
\end{equation}
where the $\lambda_{i}$ are Grassmann-odd functions of time which still
have to be determined. 
Inserting this ansatz into (\ref{fermi1-coulomb}) and
(\ref{fermi2-coulomb}) we find that the equations of motion are
satisfied as long as the coefficient functions $\lambda_{i}(t)$ obey
the condition: 
\begin{equation}
\dot{\lambda}_{i}=-\frac{1}{2^{\half}\modxz^{\thalf}}
\epsilon_{ij}\lambda_{j}.\label{lambda-equation}
\end{equation}
This first order differential equation for $\lambda_{i}$ can be solved
by:
\begin{eqnarray}
\lambda_{1}(t)&=&P\cos\omega(t)-\Pt\sin\omega(t)\label{lambda1-solution}\\
\lambda_{2}(t)&=&P\sin\omega(t)+\Pt\cos\omega(t)\label{lambda2-solution},
\end{eqnarray}
where $P$ and $\Pt$ are two Grassmann-odd constants and
\begin{equation}
\omega(t)=\int_{0}^{t}\frac{1}{2^{\half}\modxz^{\thalf}}dt'=
\frac{1}{2^{\half}}\int_{0}^{s(t)}
\frac{1}{(1\!+\!e\cosh s)^{\half}}ds\label{omega}.
\end{equation}
This integral cannot be evaluated analytically but it can be written in
terms of the standard first elliptic integral $F(s,k)$ as
\begin{displaymath}
\omega(t)=\frac{-2^{\half}i}{(e\!+\!1)^{\half}}
F\left(\ihalf s(t),\left(\frac{2e}{e\!+\!1}\right)^{\half}\right).
\end{displaymath}
We can immediately read off from (\ref{omega}) that $\omega(t)$ is a
monotonically growing function that converges to a constant as
$t\longrightarrow \pm\infty$ and that has only one zero at $t\!=\!0$. The
asymptotic behaviour of $\omega$ can be described by the formula
\begin{displaymath}
\omega_{\infty}=\lim_{t\rightarrow +\infty}\omega(t)=-
\lim_{t\rightarrow -\infty}\omega(t)=
\frac{\pi}{2e^{\half}}P_{-\half}\left(\frac{1}{e}\right),
\end{displaymath}
where $P_{\half}(k)$ is the Legendre function of first kind. 

There are two interesting limiting cases, namely $e=1$ and
$e\rightarrow\infty$. Because $P_{-1/2}(1)=1$ and $P_{-1/2}(0)$
is a finite constant we find that $\omega\rightarrow\frac{\pi}{2}$
in the first and $\omega\rightarrow 0$ in the second case. In other words,
$\omega_{\infty}$ is a monotonically falling function of $e$, taking
values in the interval $(0,\frac{\pi}{2}]$.

\bigskip
Going back to equations (\ref{lambda1-solution}) and
(\ref{lambda2-solution}) and inserting them into our ansatz
(\ref{fermion2-solution-general}) we find thus the result:
\begin{equation}
\Psi=P(\cos\omega(t)Y+\sin\omega(t)Y_{\perp})
+\Pt(-\sin\omega(t)Y+\cos\omega(t)Y_{\perp})\label{Psi-solution}
\end{equation}
which can be interpreted as follows (substituting the real components
$\Psi^{I}$, $P^{I}$, $\Pt^{I}$ for the Grassmann-valued quantities, if
necessary): In addition to motion in the $Z-Z_{\perp}$-plane the
fermionic vector can also move in the plane orthogonal to it. As in
the previous case it is rigidly attached to and corotating with two
orthogonal bosonic vectors.  The motion of these two vectors, however,
is slightly more complicated than before. They are not rigidly
connected to $Z$ and $Z_{\perp}$ in a vierbein but rather have a
time-dependent phase (unless the eccentricity tends to infinity so
that our formulae describe a freely moving particle at infinite
distance to the origin, in which case $\omega(t)\equiv0$.) 

This phase changes in a continuous fashion between $-\omega_{\infty}$ and
$\omega_{\infty}$, where $\omega_{\infty}$ depends on the
eccentricity of the bosonic solution and takes its maximum value
$\frac{\pi}{2}$ in case $e=1$, i.e. in the absence of angular momentum.  

Thus the second solution describes a circular motion orthogonal to the
first but with a time-dependent phase difference with respect to the
rigid frame provided by the bosonic vector $Z$ and its three chosen
orthogonals. We can think of this motion as parametrizing the motion
on $S_{1}$ that we mentioned above.

\bigskip
So far we have neglected the third components of $\psip$ and $\psim$,
those in the chosen direction of angular momentum, for which the
bosonic movement is trivial (i.e. $\dot{x}_{3}=U_{3}=0$). It was one
of the results from \cite{HeumannManton} that triviality of the
bosonic solution does not extend to the fermionic solution and the same
can be observed in the multidimensional case.

Since the bosonic motion takes place in the $x_{1}$-$x_{2}$-plane and
therefore $x_{3}\equiv 0$ equations (\ref{fermi1-coulomb}) and
(\ref{fermi2-coulomb}) simplify considerably: 
\begin{equation} 
\psipdt=-\frac{2^{\half}}{\modxz^{\thalf}}\psimt,\quad
\psimdt=\frac{2^{\half}}{\modxz^{\thalf}}\psipt.\label{fermion3-equation}
\end{equation}
Incidentally, up to a factor of two these are the same equations as
(\ref{lambda-equation}) and therefore have (again up to a factor of
two) the same solutions:
\begin{eqnarray}
\psipt(t)&=&O\cos 2\omega(t)-\Ot\sin 2\omega(t)\label{fermion3-solution1}\\
\psimt(t)&=&O\sin 2\omega(t)+\Ot\cos 2\omega(t),\label{fermion3-solution2}
\end{eqnarray}
where $O$ and $\Ot$ are two further Grassmann-odd constants and
$\omega(t)$ is the same function as before. The only difference 
consists of the extra factor of two and is responsible for a slight
change in the asymptotic behaviour: Whereas the maximum phase obtained
for $t\rightarrow\pm\infty$ still converges to zero for $e\rightarrow\infty$ 
implying that $\psipt=O$ and $\psimt=\Ot$ are both constants, this maximum 
phase now tends to $\pi$ rather than $\frac{\pi}{2}$ for $e=1$ meaning
that for zero angular momentum $\psipt$ and $\psimt$ describe a full circle.

\bigskip
So although there is no bosonic motion in the third direction (to
lowest order) the fermions somehow ``see'' the bosonic potential and
behave accordingly. In fact, we can think of (\ref{fermion3-equation})
as fermionic equation of motion for a one-dimensional Coulomb potential 
-- the only difference is that the strength of this potential is quadrupled 
with respect to the original problem. To see this we have to interpret
the right-hand sides of (\ref{fermion3-equation}) as 
$\mp\tilde{U}'(x)\psi_{\mp}$ where $\tilde{U}'(x)$ is given by 
\begin{displaymath}
\tilde{U}'(x)=\frac{2^{\half}}{x^{\thalf}}\:\Longrightarrow\:
\tilde{U}(x)=-\frac{2^{\thalf}}{x^{\half}}.
\end{displaymath}
Thus the imagined one-dimensional bosonic potential has to be
\begin{displaymath}
\tilde{V}(x)=\half \tilde{U}^{2}=\frac{4}{x}=4V.
\end{displaymath}
This can be seen as an explanation for the extra factor of two
appearing in the solutions (\ref{fermion3-solution1}) and
(\ref{fermion3-solution2}).

\bigskip
We want to finish this section by a short discussion of the newly
found fermionic constants $P$, $\Pt$, $O$ and $\Ot$ since there is a
major difference between them and the supercharges $Q$ and
$\Qt$. Formally we can write all four constants as conserved
quantities by simply inverting equations (\ref{fermion2-solution-general}),
(\ref{lambda1-solution}), (\ref{lambda2-solution}), (\ref{fermion3-solution1})
and (\ref{fermion3-solution2}). However, they 
cannot be seen as an original symmetry of the Lagrangian, in the
sense that they cannot be derived as Noether charges by a variation
just involving the dynamical variables. On the contrary, to derive 
those charges from the Lagrangian we have to presuppose a certain
knowledge of the system, e.g. the orientation of angular momentum, to
find the correct variation. As an example, $O$ and $\Ot$ can be
formally found as conserved quantities using the following variation:
\begin{displaymath}
\begin{array}{rclrcl}
\delta\psipt&=&-\epsilon\cos 2\omega(t),&
\delta\psimt&=&-\epsilon\sin 2\omega(t)\\
\tilde{\delta}\psipt&=&\:\:\:\epsilon\sin 2\omega(t),&
\tilde{\delta}\psimt&=&-\epsilon\cos 2\omega(t).
\end{array}
\end{displaymath}
Evidently, neither function on the right hand side is a function
of the dynamical variables only, instead we find a rather complicated
function of time involving $\omega(t)$. In addition this variation
would have to be different if we would alter the initial data of the
bosonic motion, for example, by making a different choice for the
orientiation of angular momentum. Thus $P$, $\Pt$, $O$ and $\Ot$
should be seen as ``lesser'' charges, not comparable with the two
supercharges $Q$ and $\Qt$ which are genuine Noether charges as we
have seen above.

\bigskip
There are, however, some interesting observations about the interrelation 
of these extra fermionic quantities to be made. Firstly, one can read off 
from (\ref{fermion3-solution1}) and (\ref{fermion3-solution2}) that the
product $S=i\psipt\psimt$ or, more generally,
$S=i(\Lv^{0}\cd\psip)(\Lv^{0}\cd\psim)$ is conserved, where $\Lv^{0}$
denotes the bottom layer angular momentum. This is a consequence of
the symmetry
\begin{displaymath}
\delta\psip=\epsilon(\Lv^{0}\cd\psim)\Lv^{0},\quad
\delta\psim=\epsilon(\Lv^{0}\cd\psip)\Lv^{0}.
\end{displaymath}
Since we have restricted ourselves to two Grassmann generators we can
replace $\Lv^{0}$ simply by $\Lv$ and so find a genuine symmetry of
the Lagrangian, involving only the dynamical quantities.

Moreover, between all six fermionic constants mentioned there then exists
the remarkable relation
\begin{displaymath}
i(Q\Qt+P\Pt+O\Ot)=2ER,
\end{displaymath}
where, as we have seen, the bosonic constants $R$ and $E$ are genuine
Noether charges for arbitrary Grassmann algebra.

\subsection{The top layer bosonic equation}
We finally have to solve equation (\ref{top}) for the bosonic top
layer, concentrating first on the homogeneous part of this equation,
i.e. ignoring $\av(\xvzero,\psip,\psim)$.

In the one-dimensional case there were two solutions to the analogue
of this equation, namely
\begin{displaymath}
x^{1}=c\dot{x}^{0};\quad 
x^{1}=c\dot{x}^{0}\int_{t_{0}}^{t}\frac{1}{(\dot{x}^{0})^{2}}dt'.
\end{displaymath}
Whereas the first term is still a solution to the three-dimensional
problem, the second one is not. A more fruitful generalization from the
one- to the multidimensional case is given by the idea that solutions
to the top layer equation describe variations of the bottom layer
system with the free constants of that motion. We will explain this
idea in more detail below.

\bigskip
To find an explicit form for the solutions we try the following ansatz
\begin{equation}
\xvone(t)=f_{1}(t)\,\xvzero(t)+f_{2}(t)\,\xzd(t).\label{xone-ansatz}
\end{equation}
Inserting this ansatz into (\ref{top}) and using the bottom layer
equation (\ref{bottom}) we can evaluate both sides and write them as
time-dependent linear combinations of $\xvzero$ and $\xzd$. Comparison
of the coefficients then yields the two relations
\begin{eqnarray}
\ddot{f}_{1}(t)+\frac{2}{\modxz^{3}}\dot{f}_{2}(t)&=&
-3\frac{f_{1}(t)}{\modxz^{3}}\label{aux-top-solution1}\\
2\dot{f_{1}}(t)+\ddot{f_{2}}(t)&=&0.\label{aux-top-solution2}
\end{eqnarray}
Integrating equation (\ref{aux-top-solution2}) and inserting the result
\begin{equation}
\dot{f}_{2}(t)=-2f_{1}(t)+c\label{f2}
\end{equation}
into (\ref{aux-top-solution1}) we get the following equation 
for $f_{1}(t)$:
\begin{equation}
\ddot{f}_{1}(t)=\frac{1}{\modxz^{3}}(f_{1}(t)-2c).
\end{equation}
We can recognize that the homogenous part of this equation is nothing
else than the bottom layer equation and this means that
we can write its two solutions as the two component solutions of
(\ref{bottom}) given in (\ref{boson-solution}). Since the inhomogenous
solution is just the constant $2c$ we can write
\begin{displaymath}
f_{1}(t)=C_{1}x^{0}_{1}(t)+C_{2}x^{0}_{2}(t)+2c=C_{1}(e+\cosh
s(t))-C_{2}\es\sinh s(t)+2c.
\end{displaymath}
Applying this result to (\ref{f2}) we find that
\begin{displaymath}
f_{2}(t)=-2C_{1}X_{1}(t)-2C_{2}X_{2}(t)-3ct+\tilde{c},
\end{displaymath}
where
\begin{displaymath}
\begin{array}{rcccl}
X_{1}(t)&=&\int_{0}^{t}x_{1}^{0}(t')dt'&=&\thalf es(t)+\half e \sinh
s(t)\cosh s(t) +(e^{2}-1)\sinh s(t)\\
X_{2}(t)&=&\int_{0}^{t}x_{2}^{0}(t')dt'&=&-\es(\cosh s(t)+\half
e\sinh^{2} s(t)-1).
\end{array}
\end{displaymath}
Inserting these results back into our ansatz (\ref{xone-ansatz}) we find that
\begin{eqnarray}
\xvone(t)&=&C_{1}\left(x^{0}_{1}(t)\xvzero(t)-2X_{1}(t)\xzd(t)\right)
+C_{2}\left(x^{0}_{2}(t)\xvzero(t)-2X_{2}(t)\xzd(t)\right)
\label{xone-solution}\\
&&+c\left(2\xvzero(t)-3t\xzd(t)\right)+\tilde{c}\xzd(t).\nonumber
\end{eqnarray}

There are still two solutions to (\ref{top}) missing. As all solutions
found so far lie in the plane of motion spanned by $\xvzero(t)$ and
$\xzd(t)$ we have to look now at solutions in the third direction,
i.e. for $x^{1}_{3}(t)$. The equation of motion simplifies here to
\begin{displaymath}
\ddot{x}^{1}_{3}(t)=\frac{1}{\modxz^{3}}x^{1}_{3}(t)
\end{displaymath}
since $x^{0}_{3}(t)\equiv0$. Again this is just the bottom layer
equation of motion, solved by $x^{0}_{1}(t)$ and $x^{0}_{2}(t)$.
The general solution to the homogenous part of (\ref{top}) is thus
given by (\ref{xone-solution}) and 
\begin{equation}
\xvone(t)=\tilde{C}_{1}x^{0}_{1}(t)\Lv^{0}+\tilde{C}_{2}x^{0}_{2}(t)\Lv^{0}.
\label{xone-solution2}
\end{equation}

\bigskip
When we look for an interpretation of these solutions the
analogy with the one-dimensional case will be -- as mentioned -- quite
fruitful. In one dimension the solution to the homogenous part of the
$x^{1}$-equation could be written as a sum of variations:
\begin{displaymath}
x^{1}(t)=c_{1}\frac{\delta x^{0}}{\delta t_{0}}(t)+
c_{2}\frac{\delta x^{0}}{\delta E}(t).
\end{displaymath}
The same turns out to be true in the three-dimensional case.

To start with, the term with coefficient $\tilde{c}$ in (\ref{xone-solution})  
can be clearly interpreted as variation of our bottom layer solution
with initial time $t_{0}$:
\begin{displaymath}
\xzd(t)=\frac{\delta\xvzero}{\delta t_{0}}.
\end{displaymath}

Variation with energy can also be found in the solution for $\xvone$ as can be
seen in the following way: While solution (\ref{boson-solution}) is valid
only for energy $E=\half$ we can obtain a solution for arbitrary energy
$E$ by scaling both $\xvzero$ and $t$ according to (\ref{scaling}). 
The variation of $\xvzero$ with energy is then obtained by comparing
$\xvzero_{E}(t)$ with $\xvzero_{E+dE}(t)$:
\begin{displaymath}
\frac{\delta\xvzero_{E}}{\delta E}=\frac{\partial \xvzero_{E}}{\partial E}
+\frac{\partial\xvzero_{E}}{\partial t}\frac{dt}{dE}
=-\frac{1}{2E^{2}}\xvzero+3(2E)^{\half}t_{E}\xzd_{E}.
\end{displaymath}
Evaluating this result at $E=\half$ yields
\begin{displaymath}
\left.\frac{\delta\xvzero_{E}}{\delta E}\right|_{E=\half}=-2\xvzero+3t\xzd
\end{displaymath}
leaving us with the term in (\ref{xone-solution}) with coefficient
$c$. It is quite important to realize that we have to compare two solutions of
different energy at the \textit{same} physical time $t$, i.e. we must
not change clocks and use $t_{E}$ as a parameter for the solution 
$\xvzero_{E+dE}$.

\bigskip
These two variations comprised the homogenous part of the solution
to the one-dimensional problem. For the Coulomb problem there are four
more independent ways to vary the bottom layer solution. One of these
consists of rotating the hyperbolic orbit in the plane of motion by a
small angle $\phi$ around the angular momentum axis. The variation is
thus given by
\begin{displaymath}
\frac{\delta\xvzero}{\delta\phi}=(-x^{0}_{2},x^{0}_{1},0).
\end{displaymath}
This variation coincides with a linear combination of two terms in
(\ref{xone-solution}), namely
\begin{displaymath}
\frac{\delta\xvzero}{\delta\phi}=-\frac{e}{e^{2}-1}\left[
\left(x^{0}_{2}\xvzero-2X_{2}\xzd\right)
+\es\left(e^\half+e^{-\half}\right)^{2}\xzd\right].
\end{displaymath}
We can thus interpret the term with coefficient $C_{2}$ in
(\ref{xone-solution}) as a combination of rotation in the plane of
motion and variation of the initial time parameter.

\bigskip
Another way to vary $\xvzero$ is to change the eccentricity of the
hyperbola which is explicitly included as a parameter in the solution
(\ref{boson-solution}). We find
\begin{displaymath}
\frac{\delta \xvzero}{\delta e}=\frac{\partial \xvzero}{\partial e}
+\frac{\partial \xvzero}{\partial s}\frac{\partial s}{\partial e}
=\left(1-\frac{x^{0}_{2}\dot{x}^{0}_{1}}{L^{0}_{3}},
\frac{x^{0}_{1}\dot{x}^{0}_{1}}{L^{0}_{3}},0\right),
\end{displaymath}
where we have used that
\begin{displaymath}
\frac{\partial s}{\partial e}=
\left.\left(\frac{\partial e}{\partial s}\right)^{-1}\right|_{t=const}
=-\frac{\sinh s}{1+e\cosh s}.
\end{displaymath}
We again find this variation as a linear combination of two terms in
(\ref{xone-solution}), namely:
\begin{displaymath}
\frac{\delta \xvzero}{\delta e}=-\frac{1}{e^{2}\!-\!1}
\left[\left(x^{0}_{1}\xvzero-2X_{1}\xzd\right)
-e\left(2\xvzero-3t\xzd\right)\right],
\end{displaymath}
and so the the $C_{1}$-term in (\ref{xone-solution}) describes a
combination of variation with eccentricity and energy.

\bigskip
It is now to be expected that the two remaining terms of the
homogenous $\xvone$-solution are connected to a tilt of the plane of
motion or, equivalently, to a change in the orientation of angular
momentum $\Lv^{0}$:
\begin{displaymath}
\delta\Lv^{0}=\delta L_{1}^{0}\eone+\delta L_{2}^{0}\etwo,
\end{displaymath}
here $\eone$ and $\etwo$ represent unit vectors in the respective
directions. We find that
\begin{displaymath}
\frac{\delta\xvzero}{\delta L_{1}^{0}}=x^{0}_{1}\ethree,\quad
\frac{\delta\xvzero}{\delta L_{2}^{0}}=x^{0}_{2}\ethree,
\end{displaymath}
which are evidently proportional to the two terms in
(\ref{xone-solution2}) since $\Lv^{0}$ is oriented in the third direction
in our particular solution to the bottom layer system.

\bigskip
Thus we have now derived and found an interpretation for all the terms
of the homogeneous solution to the top layer system. As was the case
for one-dimensional systems this solution describes all the possible
variations of the bottom layer with all possible parameters describing
this layer. In our case we need six parameters to describe the motion
of the original Coulomb problem (initial time, energy, eccentricity,
orientation of major axis and two parameters describing the plane of
motion) and correspondingly there are six independent solutions to the
top layer.

\bigskip
The last step is now to find one particular solution to the
inhomogenous equation (\ref{top}). The general idea is to take linear
combinations of the inhomogeneous terms $a_{i}$, integrate them over
time and multiply them by one of the solutions $\xv^{1}_{hom}$ of the
homogeneous equation, then sum over all solutions:
\begin{displaymath}
\xv_{inhom}^{1}=\sum_{\alpha=1}^{n}\xv^{1}_{hom,\alpha}
\int_{0}^{t}f_{\alpha,i}(t')a_{i}(t')dt'.
\end{displaymath}
We determine the coefficient functions $f_{\alpha,i}$ by two
conditions: Firstly, in taking the time derivative we want to get
\begin{displaymath}
\dot{\xv}^{1}_{inhom}=\sum_{\alpha=1}^{n}\dot{\xv}^{1}_{hom,\alpha}
\int_{0}^{t}f_{\alpha,i}(t')a_{i}(t')dt',
\end{displaymath}
i.e. all the terms involving derivatives of the integrals must cancel out
each other. Then for the second time derivative we arrive at
\begin{displaymath}
\ddot{\xv}^{1}_{inhom}=\sum_{\alpha=1}^{n}\ddot{\xv}^{1}_{hom,\alpha}
\int_{0}^{t}f_{\alpha,i}(t')a_{i}(t')dt'+
\left(\sum_{\alpha=1}^{n}\dot{\xv}^{1}_{hom,\alpha}f_{\alpha,i}(t)\right)
a_{i}(t).
\end{displaymath}
Because the $\xv^{1}_{hom,\alpha}$-terms satisfy the homogeneous equation 
the first sum will give the homogeneous part of (\ref{top}). For the
next term to give the inhomogeneous part we need the second condition 
that the sum in parenthesis is equal to the unit vector $\ev^{i}$.

\bigskip
We will now demonstrate this procedure: For simplicity we begin with
the third component of $\xvone$. Since it decouples from the other
components we can safely assume $f_{\alpha,1}=f_{\alpha,2}=0$,
i.e. writing $f_{\alpha}$ instead of $f_{\alpha,3}$ we can make the ansatz
\begin{equation}
x^{1}_{inhom,3}=x^{0}_{1}\int_{0}^{t}f_{1}(t')a_{3}(t')dt'
+ x^{0}_{2}\int_{0}^{t}f_{2}(t')a_{3}(t')dt'.\label{inhomogen-solution1}
\end{equation}
Our first condition then yields $x^{0}_{1}(t)f_{1}(t)a_{3}(t)
+x^{0}_{2}(t)f_{2}(t)a_{3}(t)=0$ and thus we can set
$f_{1}=cx^{0}_{2}$ and $f_{2}=-cx^{0}_{1}$. Then we find
\begin{eqnarray*}
\ddot{x}^{1}_{inhom,3}&=&c\left(
\ddot{x}^{0}_{1}\int_{0}^{t}x^{0}_{2}(t')a_{3}(t')dt'
-\ddot{x}^{0}_{2}\int_{0}^{t}x^{0}_{1}(t')a_{3}(t')dt'\right)
+c(\dot{x}^{0}_{1}x^{0}_{2}-\dot{x}^{0}_{2}x^{0}_{1})a_{3}(t)\\
&=&\frac{1}{\modxz^{3}}c\left( x^{0}_{1}\int_{0}^{t}x^{0}_{2}(t')a_{3}(t')dt'
- x^{0}_{2}\int_{0}^{t}x^{0}_{1}(t')a_{3}(t')dt'\right)
-cL^{0}_{3}a_{3}(t)
\end{eqnarray*}
and so by equating $c=-\frac{1}{L^{0}_{3}}$ we end up with the
required result, namely
\begin{displaymath}
\ddot{x}^{1}_{inhom,3}=\frac{1}{\modxz^{3}}x^{1}_{3}+a_{3}(t).
\end{displaymath}
Since the equations for the first and second components are coupled we
need all four remaining homogenous solutions to give the correct
inhomogeneous term. We just write down the result:
\begin{eqnarray}
\xvone_{inhom}&=&\frac{1}{L^{0}_{3}}\left[\xzd\int_{0}^{t}
a_{1}\left(2\frac{X_{1}}{L^{0}_{3}}\dot{x}^{0}_{2}x^{0}_{2}
+2\frac{X_{2}}{L^{0}_{3}}(L^{0}_{3}-\dot{x}^{0}_{1}x^{0}_{2})-3tx^{0}_{2}
\right)\right.\label{inhomogen-solution2}\\
&&\quad\quad\quad\quad
+a_{2}\left(2\frac{X_{2}}{L^{0}_{3}}\dot{x}^{0}_{1}x^{0}_{1}
-2\frac{X_{1}}{L^{0}_{3}}(L^{0}_{3}+\dot{x}^{0}_{2}x^{0}_{1})+3tx^{0}_{1}
\right)dt'\nonumber\\
&&+(2\xvzero-3t\xzd)\int_{0}^{t}\left(
-a_{1}x^{0}_{2}+a_{2}x^{0}_{1}\right)dt'\nonumber\\
&&+(x^{0}_{1}\xvzero-2X_{1}\xzd)\int_{0}^{t}\left(
a_{1}\frac{\dot{x}^{0}_{2}x^{0}_{2}}{L^{0}_{3}}-
a_{2}(1+\frac{\dot{x}^{0}_{2}x^{0}_{1}}{L^{0}_{3}})\right)dt'\nonumber\\
&&\left.+(x^{0}_{2}\xvzero-2X_{2}\xzd)\int_{0}^{t}\left(
a_{1}(1-\frac{\dot{x}^{0}_{1}x^{0}_{2}}{L^{0}_{3}})+
a_{2}\frac{\dot{x}^{0}_{1}x^{0}_{1}}{L^{0}_{3}}\right)dt'\right].\nonumber
\end{eqnarray}
Because we know both the bottom layer bosonic and the fermionic
quantities contained in $\av$ we can now in principle insert these
into the integrals and evaluate the solutions as explicit functions of
time. However, since the knowledge of these functions does not provide
new insights, we refrain from writing them down here.

\bigskip
In summary, the complete solution to the top layer bosonic equation is
given by the sum of the homogeneous solutions (\ref{xone-solution})
and (\ref{xone-solution2}) and the inhomogeneous solution given by
(\ref{inhomogen-solution1}) and (\ref{inhomogen-solution2}).

\section{Discussion}\label{discussion}

Our aim in this paper has been not only to analyze and solve a
fascinating problem in supersymmetric classical mechanics but also to
stress the relations between supersymmetric mechanical models in one
and more than one dimensions. An example is that the fermionic
variables should really be thought of as components of a
$2n$-dimensional vector $\Psi$ which moves on a $(2n\!-\!1)$-sphere. In the
one-dimensional case this means that all fermionic vectors move on
circles -- which is consistent with our findings in
\cite{HeumannManton}. Furthermore, the motion is coupled to the motion
of a purely bosonic vector, made of the dynamical variables $\xd$
and $\U$, by the supercharges. While these completely determine the
fermionic motion in one dimension we have seen that the situation in
higher dimensions is more complicated and the supercharges only
prescribe a geometric subspace for the true solution. Motion on this
subspace is then no longer characterized by the rigid connection of
bosonic and fermionic motion found in the one-dimensional case but
exhibits a rather interesting time-dependent phase.

\bigskip
We have shown that for rotationally invariant scalar potentials there
will always be a conserved angular momentum so that to lowest
order bosonic motion takes place in a plane. It is not unreasonable to
conjecture that for every direction orthogonal to that plane we will
have to solve a one-dimensional problem for the fermionic quantities 
corresponding to that direction -- as was the case for the Coulomb
problem. This leads us to another generalization from the
one-dimensional case, namely that trivial solutions to the bosonic
equations of motion do not imply that the corresponding fermionic
equations can only be solved trivially. In fact, for our particular
problem we found that the third components of the fermionic quantities
still ``felt'' the presence of the bosonic motion in the plane
orthogonal to them.

\bigskip
One of the most fruitful ideas for the higher bosonic layers which can
be applied in any dimension turned out to be that solutions for these
layers are variations of the solutions of the classical problem. While
in one dimension every solution can be conveniently described by the
energy and some choice for the initial time, in higher dimensions more
of these parameters are needed -- in case of the Coulomb problem we have
conveniently related them to the elements describing the classical orbit. 
So the view taken in \cite{HeumannManton} for the one-dimensional
case, namely that supersymmetric dynamics captures information over a
whole range of \textit{energies} of the system, can be confirmed in a
more general sense for our higher dimensional model: The bosonic (but
Grassmann-valued) part of the solution describes not only one
particular classical solution (by its component of order zero) but
also includes all possible variations of it with all possible free parameters
or, in other words, a bosonic solution corresponds not just to a point in
the real parameter space but to something like a fuzzy subset.

For reasons of simplicity we have described in this paper solutions
for a Grassmann algebra with two generators. From our experience with
the one-dimensional model and the clear links that it exhibits to the
three-dimensional system we have studied in this paper, we conjecture
that the picture for a larger number of generators is similar to that
in the one-dimensional case. There we were able to show that solutions
to the fermionic equations corresponding to three generators consist
of two parts: One that looks like the first order solutions themselves
(i.e. movement on a circle in one dimension) and one that consists of
the \textit{variations} of these solutions with energy and initial
time. Similarly, the bosonic equations for four generators are solved
by the first and second variations of the lowest order bosonic solutions
\textit{and} first variations of the interaction term.

It is therefore not unreasonable to assume that the general picture in three
dimensions is not much different and that the higher order equations
of motion can be solved by a suitable combination of (first and higher
order) variation terms with respect to the free parameters of the 
physical model in question.

\section*{Appendix: Runge-Lenz-Vector}\label{appendix}
In this section we want to outline a proof that the supersymmetric
Coulomb problem discussed in this paper does not admit a conserved Runge-Lenz
vector. In the classical problem this conserved quantity cannot be
constructed as a Noether charge but arises as a consequence of the
hidden $SO(3,1)$-symmetry of the problem, explicitly recognizable in our
construction of the bottom layer solution to the equations of motion
in section \ref{sectionbottom}. This symmetry must be broken in the
supersymmetric case.

\bigskip
We start with the assumption that there is a conserved Runge-Lenz
vector in the supersymmetric problem. Setting all fermionic variables
to zero (for our choice of Grassmann algebra this can be accomplished
by choosing $Q\!=\!\Qt\!=\!P\!=\!\Pt\!=\!O\!=\!\Ot\!=\!0$) we expect to
get the classical result for this vector, namely
\begin{equation}
\Vv=[\Lv,\pv]-\xh,\label{Runge-Lenz}
\end{equation}
where $\Lv$ is given just by its bosonic part $[\xv,\pv]$. (We will
use $\pv\equiv\xd$ for this section). 

Notice that the Hamiltonian and the angular momentum vector also have
this property. Of course, here all quantities are Grassmann-even rather
than real, but this makes no difference and, of course, (\ref{Runge-Lenz}) 
is a conserved quantity if we set all fermionic terms in the Lagrangian 
(\ref{Lagrangian}) to zero.

\bigskip
Returning now to the general case with non-trivial fermionic variables
we take the time-derivative of (\ref{Runge-Lenz}) and find (still taking
the classical result for $\Lv$):
\begin{eqnarray*}
\frac{d}{dt}\Vv\!&=&\!-\thalf i\frac{2^{\half}}{\modx^{\thalf}}\left[
\pv\!\left((\psip\!\cd\psim)-\thalf(\psip\!\cd\xh)(\psim\!\cd\xh)\right)+
\psip(\psim\!\cd\xh)(\pv\cd\!\xh)-\psim(\psip\!\cd\xh)(\pv\cd\!\xh)
\right.\\
&&\left.\!-\xh\!\left(2(\psip\!\cd\pv)(\psim\!\cd\xh)\!+\!2(\psip\!\cd\xh)
(\psim\!\cd\pv)\!+\!(\psip\!\cd\psim)(\pv\cd\!\xh)
\!-\!\frac{7}{2}(\psip\!\cd\xh)(\psim\!\cd\xh)(\pv\cd\!\xh)
\right)\right].
\end{eqnarray*}
The result is clearly non-zero but then we can expect the supersymmetric
version of the Runge-Lenz vector to have an additional fermionic piece. 
However, the derivative of this piece should yield exactly the same
result as above with an extra minus sign so that the derivative of the total
will be zero and hence $\Vv$ a conserved quantity. We will now try to
construct this fermionic piece.

\bigskip
We start by writing our result for the bosonic part of $\Vv$ in
components, so that the linearity in $\pv$, $\psip$ and $\psim$ can be
seen in a more explicit way:
\begin{eqnarray}
\frac{d}{dt}V_{RL}^{k}&=&-\thalf i\frac{2^{\half}}{\modx^{\thalf}}\left(
\delta^{ij}\delta^{kl}\!-\!\thalf\delta^{kl}\xhc^{i}\xhc^{j}\!+\!
\delta^{ik}\xhc^{j}\xhc^{l}\!+\!\delta^{jk}\xhc^{i}\xhc^{l}\right.\nonumber\\
&&\left.-2\delta^{il}\xhc^{j}\xhc^{k}\!-\!2\delta^{jl}\xhc^{i}\xhc^{k}\!-\!
\delta^{ij}\xhc^{k}\xhc^{l}\!+\!\frac{7}{2}\xhc^{i}\xhc^{j}\xhc^{k}\xhc^{l}
\right)\psi_{+}^{i}\psi_{-}^{j}p^{l}.\label{RV-decomposition}
\end{eqnarray}
We now try to assemble all terms whose derivative could possibly
contribute to this result. Fortunately, there are restrictions as to
which terms to consider that shorten the list of expressions to look at. 

Firstly, we need only to analyze terms quadratic in the fermionic variables: 
Since $\Vv$ is even their number has to be divisible by two, all terms
containing no fermionic quantities at all can be assumed to be contained 
already in (\ref{Runge-Lenz}) and if there were four or more fermionic
variables in $\Vv$ the time derivative would still contain the same
number of these variables in contradiction to (\ref{RV-decomposition}). 
Furthermore since our system is symmetric under the exchange
$\psip\rightarrow -\psim, \psim\rightarrow\psip$ we have to include
only terms subject to this symmetry. This leaves us with multiples of 
the terms 
\begin{displaymath}
\begin{array}{l}
\mbox{A)}\quad i\psi_{+}^{i}\psi_{-}^{j}\\
\mbox{B)}\quad i\psi_{+}^{i}\psi_{+}^{j}+i\psi_{-}^{i}\psi_{-}^{j}.
\end{array}
\end{displaymath}
When we now classify all terms according to the type of their
fermionic part and the even or odd number of $p^{l}$-variables
involved in each one of them, then the time derivative can be seen as
acting on these classes as follows:
\begin{center}
\begin{tabular}{llllll}
&&$\frac{d}{dt}$&&&\\
I)&type A,  $p^{l}$ even&$\rightarrow$&type A,  $p^{l}$ odd
&+&type B,  $p^{l}$ even\\
II)&type A,  $p^{l}$ odd&$\rightarrow$&type A,  $p^{l}$ even
&+&type B,  $p^{l}$ odd\\
III)&type B,  $p^{l}$ even&$\rightarrow$&type A,  $p^{l}$ even
&+&type B,  $p^{l}$ odd\\
IV)&type B,  $p^{l}$ odd&$\rightarrow$&type A,  $p^{l}$ odd
&+&type B,  $p^{l}$ even.
\end{tabular}
\end{center}
Here $p^{l}$ even/odd means the number of $p^{l}$-variables contained
in a term is even or odd, respectively. 

\bigskip
Since we want to derive a term of type A with an \textit{odd} number
of $p^{l}$'s, namely just one, we can read off from the summary above
that we need not consider combinations of type II or III, since their
time derivatives do not yield such a term nor can they compensate the
B-terms that will arise from combinations I or IV. This means we only
have to consider these latter combinations. However, even those are
not unproblematic: As mentioned, they give rise to fermionic 
terms of type B which we do not want in our derivative, so we must
hope that we can cancel these terms against each other.

\bigskip
In the next step we make -- preliminarily -- the assumption that the
number of $p^{l}$-variables involved is either zero or one. We will
later see that allowing for higher powers of $p^{l}$ will not change
anything.

Up to this point the allowed combinations left are 
\begin{displaymath}
\mbox{a)}i\psi_{+}^{i}\psi_{-}^{j},\quad
\mbox{b)}(i\psi_{+}^{i}\psi_{+}^{j}+i\psi_{-}^{i}\psi_{-}^{j})p^{k}.
\end{displaymath}
We now have to contract these with the remaining natural tensors, namely
$\delta^{ij}$ and $\xhc^{i}$, to leave only one free index. For terms
of type a) we have the following options:
$\xhc^{i}\xhc^{j}\xhc^{k}$,$\,\xhc^{i}\delta^{jk}$,
$\xhc^{j}\delta^{ik}$,$\,\xhc^{k}\delta^{ij}$.

For terms of type b) we have some extra restrictions: First, since
they are antisymmetric under the exchange $i\leftrightarrow j$ we need
not to consider $\delta^{ij}$-terms that contract the two fermionic
variables. Second, if we include any $\xhc$-term the time-derivative 
will automatically yield an additional $p^{l}$-term so that the result
would be quadratic in the $p^{l}$'s -- a case which we have excluded
above. Therefore the only option left for combination b) is
$\delta^{jk}\delta^{il}$.

We still have to determine the prefactor of each term. It
can be derived by dimensional arguments. As can be read off from the
Coulomb-Hamiltonian (\ref{Coulomb-Hamiltonian}) the dimension of
energy is $\mbox{length}^{-1}$. From this we can calculate the
dimension of every other dynamical quantity: 
\begin{displaymath}
\begin{array}{|l|l|l|l|l|}\hline
\xv&\pv&\psip&\psim&\Vv\\\hline
\mbox{length}&\mbox{length}^{-1/2}&\mbox{length}^{1/4}
&\mbox{length}^{1/4}&1\\\hline
\end{array}
\end{displaymath}

Since the bosonic part of the Runge-Lenz vector is dimensionless we
have to multiply each candidate term for the fermionic part by the
appropriate factor of $r\equiv\modx$ to make it dimensionless,
too. We end up with the list in the first column of the following
two tables. These tables describe the action of the time-derivative on
each candidate term. In the top rows we have specified all tensorial 
combinations that can possibly be generated by applying the time-derivative 
to the candidate terms in the first column. The table itself then
specifies the correct numerical prefactors of each tensor in the
time-derivative of a particular candidate term. For example, one can
read off from the table entries for the third term that
\begin{eqnarray*}
\frac{d}{dt}\left(\frac{1}{r^{1/2}}\!\xhc^{j}\!\delta^{ik}
i\psi_{+}^{i}\psi_{-}^{j}\right)
&=&\left(1\cd\delta^{ik}\delta^{jl}
+\left(-\thalf\right)\cd\delta^{ik}\xhc^{j}\xhc^{l}\right)
r^{-\thalf}p^{l}i\psi_{+}^{i}\psi_{-}^{j}\\
&&+\half\cd\sqrt{2}r^{-2}\delta^{jk}\xhc^{i}\!i\psi_{+}^{i}\psi_{+}^{j}+
1\cd\sqrt{2}r^{-2}\delta^{jk}\xhc^{i}\!i\psi_{-}^{i}\psi_{-}^{j}.
\end{eqnarray*}
For comparison we have added the bosonic part of the Runge-Lenz vector
and the tensor decomposition of its time-derivative in the last row.
The result is the following:

\begin{displaymath}
\begin{array}{|l||c|c|c|c|c|c|c|c|c|c|}\hline
&\multicolumn{10}{|c|}
{r^{-3\!/2\!}p^{l}i\psi_{+}^{i}\psi_{-}^{j}}
\\\cline{2-11}
&\!\delta^{il}\!\delta^{jk}&\!\delta^{ik}\!\delta^{jl}&
\!\delta^{ij}\!\delta^{kl}&\!\delta^{il}\!\xhc^{j}\!\xhc^{k}&
\!\delta^{jl}\!\xhc^{i}\!\xhc^{k}&\!\delta^{jk}\!\xhc^{i}\!\xhc^{l}&
\!\delta^{ik}\!\xhc^{j}\!\xhc^{l}&\!\delta^{kl}\!\xhc^{i}\!\xhc^{j}&
\!\delta^{ij}\!\xhc^{k}\!\xhc^{l}&\!\xhc^{i}\!\xhc^{j}\!\xhc^{k}\!\xhc^{l}
\\\hline\hline
1)\frac{1}{r^{1\!/\!2}}\!\xhc^{i}\!\xhc^{j}\!\xhc^{k}
i\psi_{+}^{i}\psi_{-}^{j}&
0&0&0&1&1&0&0&1&0&-\frac{7}{2}\\\hline
2)\frac{1}{r^{1\!/\!2}}\!\xhc^{i}\!\delta^{jk}i\psi_{+}^{i}\psi_{-}^{j}&
1&0&0&0&0&-\thalf&0&0&0&0\\\hline
3)\frac{1}{r^{1\!/\!2}}\!\xhc^{j}\!\delta^{ik}i\psi_{+}^{i}\psi_{-}^{j}&
0&1&0&0&0&0&-\thalf&0&0&0\\\hline
4)\frac{1}{r^{1\!/\!2}}\!\xhc^{k}\!\delta^{ij}i\psi_{+}^{i}\psi_{-}^{j}&
0&0&1&0&0&0&0&0&-\thalf&0\\\hline
5)\delta^{il}\!\delta^{jk}p^{l}(i\psi_{+}^{i}\psi_{+}^{j}&
0&0&0&\thalf\!\sqrt{2}&\thalf\!\sqrt{2}&-\thalf\!\sqrt{2}&-\thalf\!\sqrt{2}&
0&0&0\\
\quad+i\psi_{-}^{i}\psi_{-}^{j})&&&&&&&&&&\\\hline
\mbox{\boldmath$V$}_{RL,bos}
&0&0&-\thalf\!\sqrt{2}&3\!\sqrt{2}&3\!\sqrt{2}&-\thalf\!\sqrt{2}
&-\thalf\!\sqrt{2}&\frac{9}{4}\!\sqrt{2}&\thalf\!\sqrt{2}&
-\frac{21}{4}\!\sqrt{2}
\\\hline\hline
&\multicolumn{5}{|c|}{\sqrt{2}r^{-2}\delta^{jk}\xhc^{i}\!
i\psi_{+}^{i}\psi_{+}^{j}}
&\multicolumn{5}{|c|}{\sqrt{2}r^{-2}\delta^{jk}\xhc^{i}\!
i\psi_{-}^{i}\psi_{-}^{j}}
\\\hline\hline
1)\frac{1}{r^{1\!/\!2}}\!\xhc^{i}\!\xhc^{j}\!\xhc^{k}i\psi_{+}^{i}\psi_{-}^{j}
&\multicolumn{5}{|c|}{0}&\multicolumn{5}{|c|}{0}\\\hline
2)\frac{1}{r^{1\!/\!2}}\xhc^{i}\delta^{jk}i\psi_{+}^{i}\psi_{-}^{j}
&\multicolumn{5}{|c|}{1}&\multicolumn{5}{|c|}{\half}\\\hline
3)\frac{1}{r^{1\!/\!2}}\xhc^{j}\delta^{ik}i\psi_{+}^{i}\psi_{-}^{j}
&\multicolumn{5}{|c|}{\half}&\multicolumn{5}{|c|}{1}\\\hline
4)\frac{1}{r^{1\!/\!2}}\xhc^{k}\delta^{ij}i\psi_{+}^{i}\psi_{-}^{j}
&\multicolumn{5}{|c|}{0}&\multicolumn{5}{|c|}{0}\\\hline
5)\delta^{il}\delta^{jk}p^{l}(i\psi_{+}^{i}\psi_{+}^{j}
&\multicolumn{5}{|c|}{\half\sqrt{2}}&\multicolumn{5}{|c|}{\half\sqrt{2}}\\
\quad+i\psi_{-}^{i}\psi_{-}^{j})
&\multicolumn{5}{|c|}{}&\multicolumn{5}{|c|}{}\\\hline
\mbox{\boldmath$V$}_{RL,bos}
&\multicolumn{5}{|c|}{0}&\multicolumn{5}{|c|}{0}\\\hline
\end{array}
\end{displaymath}

We now have to choose appropriate coefficients for the candidate terms
such that their sum equals the bosonic part of $\Vv$ (with an extra
minus sign). Looking at the coefficients of $\delta^{il}\delta^{jk}$ and 
$\delta^{ik}\delta^{jl}$ we can immediately see that terms 2) and 3)
cannot occur since these tensorial combinations
are not part of the bosonic Runge-Lenz vector and cannot be
compensated by any other term. Looking then at the second table we
can conclude that term (5) cannot occur either -- since it contributes
a non-zero factor to both columns whereas $\Vv$ does not (and the
terms 2) and 3) have to be zero as deduced above).  

The column for $\delta^{ij}\delta^{kl}$ tells us that the coefficient
of the fourth term has to be $\thalf\sqrt{2}$ -- in contradiction to
the column for $\delta^{ij}\xhc^{k}\xhc^{l}$ which says that the same
coefficient has to be $\sqrt{2}$. Similar contradictions arise for our
first candidate term when we compare the columns for
$\delta^{il}\xhc^{j}\xhc^{k}$, $\delta^{jl}\xhc^{i}\xhc^{k}$,
$\delta^{kl}\xhc^{i}\xhc^{j}$ and $\xhc^{i}\xhc^{j}\xhc^{k}\xhc^{l}$.
This leaves us no other option than to conclude that none of our
candidate terms can be added to the bosonic part of the Runge-Lenz
vector to achieve a zero derivative.

\bigskip
One loophole remains. We have excluded further terms of type
b) on the grounds that they yield terms quadratic in momentum. We might
hope, however, that in the right combination these quadratic terms
can cancel each other. There are two possibilities, which can be found
in the first column of the following table; all other tensorial
combinations either yield zero or are up to minus sign identical to
the two choices below via the index exchange $i\leftrightarrow j$.
On taking the time derivative of both terms the crucial issue is
whether the terms quadratic in $p^{l}$ can be cancelled against each
other, so only quadratic components are shown here:
\begin{displaymath}
\begin{array}{|l|c|c|c|c|c|}\hline
&\multicolumn{5}{|c|}{\frac{1}{r}p^{l}p^{m}(i\psi_{+}^{i}\psi_{+}^{j}
\!+\!i\psi_{-}^{i}\psi_{-}^{j})}\\\cline{2-6}
&\xhc^{i}\delta^{jm}\delta^{kl}&\xhc^{i}\delta^{jk}\delta^{lm}&
\xhc^{l}\delta^{im}\delta^{jk}&\xhc^{i}\xhc^{k}\xhc^{l}\delta^{jm}&
\xhc^{i}\xhc^{l}\xhc^{m}\delta^{jk}\\\hline
\xhc^{i}\xhc^{k}\delta^{jl}p^{l}(i\psi_{+}^{i}\psi_{+}^{j}
\!+\!i\psi_{-}^{i}\psi_{-}^{j})&1&0&0&-2&0\\\hline
\xhc^{i}\xhc^{l}\delta^{jk}p^{l}(i\psi_{+}^{i}\psi_{+}^{j}
\!+\!i\psi_{-}^{i}\psi_{-}^{j})&0&1&1&0&-2\\\hline
\end{array}
\end{displaymath}
Evidently, the quadratic terms do not cancel each other. We conclude
that no term at most linear in momentum exists that could be
added to $\mbox{\boldmath$V$}_{RL,bos}$ to give a conserved quantity
in the supersymmetric model.

We can of course ask the question whether the fermionic part of the
Runge-Lenz vector could be quadratic or of some higher power in
$p^{k}$. We will now show that the answer to this question is no. The
reason for this can be understood by looking at a general term like
\begin{equation}
A^{ijk_{1}\dots k_{m}l}f(r)p^{k_{1}}\dots p^{k_{m}}i\psi_{+}^{i}\psi_{-}^{j},
\label{candidate-Runge-Lenz}
\end{equation}
where $A^{ijk_{1}\dots k_{m}l}$ ($l$ is a free index) is a tensor
constructed from two types of building blocks, $\delta^{pq}$ and
$\xhc^{p}$, and $f(r)$ is some well-determined power of $r$. Since the
tensor is of type A we can infer that the number $m$ of momentum
variables has to be even. Furthermore, to keep (\ref{candidate-Runge-Lenz}) 
dimensionless $f(r)$ has to be $r^{\half(m-1)}$. Clearly for every
$m\ge2$ $f(r)$ is a non-trivial function of $r$. This in turn means
that the time-derivative in acting on $f(r)$ will produce an extra
$p^{k}$ variable.  

\bigskip
There are only two ways to solve this problem: Either all the terms in
the time derivative of (\ref{candidate-Runge-Lenz}) containing extra momentum 
variables cancel each other -- or we have to make use of yet higher powers
of $p^{k}$ in $\mbox{\boldmath$V$}_{RL,ferm.}$ to compensate. 
However, those very terms will by the same principle give rise to
another extra power of $p^{k}$ in the derivative and so on. For this
series to stop at some point we must hope that all terms in the
derivative of $A^{ijk_{1}\dots k_{m}l}$ with $m\!+\!1$ momentum
variables cancel each other. We find that
\begin{eqnarray*}
\lefteqn{\frac{d}{dt}\left(A^{ijk_{1}\dots k_{m}l}r^{\half(m-1)}p^{k_{1}}
\dots p^{k_{m}} i\psi_{+}^{i}\psi_{-}^{j}\right)=
\frac{d}{dt}(A^{ijk_{1}\!\dots\! k_{m}l})
r^{\half(m\!-\!1)}p^{k_{1}}\!\dots\! p^{k_{m}}i\psi_{+}^{i}\psi_{-}^{j}}\\
&&+A^{ijk_{1}\!\dots\!k_{m}l}\half(m\!-\!1)r^{\half(m-3)}\xhc^{k_{m+1}}
p^{k_{m+1}}p^{k_{1}}\!\dots\! p^{k_{m}}i\psi_{+}^{i}\psi_{-}^{j}+\dots,
\end{eqnarray*}
where the dots indicate further terms containing less than $m+1$
momentum variables, in which we are not interested here.

As mentioned above $A^{ijk_{1}\dots k_{m}l}$ consists of a sum of
tensors built from $\delta^{pq}$ and $\xhc^{p}$ terms. While the time
derivative does not act on the first type of term, it acts on the second:
\begin{displaymath}
\frac{d}{dt}\xhc^{p}=\frac{1}{r}(\delta^{pl}-\xhc^{p}\xhc^{l})p^{l}.
\end{displaymath}
Thus we can write
\begin{displaymath}
\frac{d}{dt}A^{ijk_{1}\dots k_{m}l}=\frac{1}{r}
\tilde{A}^{ijk_{1}\dots k_{m}k_{m+1}l} p^{k_{m\!+\!1}},
\end{displaymath}
where $\tilde{A}^{ijk_{1}\dots k_{m}k_{m\!+\!1}l}$ is a new tensor derived 
from $A^{ijk_{1}\dots k_{m}l}$ in the following way, which can be seen as an 
algebraic version of the Leibniz rule: We replace one $\xhc^{p}$ tensor in 
$A^{ijk_{1}\dots k_{m}l}$ with $(\delta^{pk_{m+1}}-\xhc^{p}\xhc^{k_{m+1}})$, 
repeat this step for all other $\xhc^{p}$ tensors contained in 
$A^{ijk_{1}\dots k_{m}l}$ -- replacing always one tensor at a time --
and then sum up every contribution. Therefore
\begin{eqnarray*}
\frac{d}{dt}\left(A^{ijk_{1}\dots k_{m}l}r^{\half(m-1)}p^{k_{1}}\dots p^{k_{m}}
i\psi_{+}^{i}\psi_{-}^{j}\right)&=&\left(\tilde{A}^{ijk_{1}\dots k_{m}k_{m+1}l}
+\half(m\!-\!1)A^{ijk_{1}\dots k_{m}l}\xhc^{k_{m+1}}\right)\\
&&\quad r^{\half(m-3)}p^{k_{1}}\!\dots\! p^{k_{m}}p^{k_{m+1}}
i\psi_{+}^{i}\psi_{-}^{j}+\dots
\end{eqnarray*}
So, if we want a cancellation of all terms containing $m+1$ momentum
variables, we need:
\begin{equation}
\tilde{A}^{ijk_{1}\dots k_{m}k_{m+1}l}+\half(m\!-\!1)A^{ijk_{1}\dots k_{m}l}
\xhc^{k_{m+1}}=0.
\label{RL-condition}
\end{equation}
It remains to show that this equation cannot be satisfied. Starting with
\begin{equation}
A_{0}^{ijk_{1}\dots k_{m}l}=
\xhc^{i}\xhc^{j}\xhc^{k_{1}}\dots\xhc^{k_{m}}\xhc^{l}
\label{RL-test}
\end{equation}
we find
\begin{displaymath}
\tilde{A}_{0}^{ijk_{1}\dots k_{m}k_{m+1}l}=
-(m\!+\!3)\xhc^{i}\xhc^{j}\xhc^{k_{1}}\dots
\xhc^{k_{m}}\xhc^{k_{m+1}}\xhc^{l}
+\delta^{ik_{m+1}}\xhc^{j}\xhc^{k_{1}}\dots\xhc^{k_{m}}\xhc^{l}+\dots+
\xhc^{i}\xhc^{j}\xhc^{k_{1}}\dots\xhc^{k_{m}}\delta^{lk_{m+1}}.
\end{displaymath}
Inserting these results into (\ref{RL-condition}) we find the necessary
condition $-\half(m\!+\!7)=0$, i.e. $m\!=\!-7$ which is a clear
contradiction, so a term like (\ref{RL-test}) cannot occur. Replacing
two $\xhc^{k}$ terms by a $\delta$-tensor we arrive at
\begin{equation}
A_{1}^{ijk_{1}\dots k_{m}l}=\xhc^{i}\xhc^{j}\xhc^{k_{1}}\dots\xhc^{k_{m-1}}
\delta^{k_{m}l}
\end{equation}
to name but one possible choice. The corresponding $\tilde{A}$ is then given by
\begin{displaymath}
\tilde{A}_{1}^{ijk_{1}\dots k_{m}k_{m+1}l}=
-(m\!+\!1)\xhc^{i}\xhc^{j}\xhc^{k_{1}}\dots
\xhc^{k_{m-1}}\delta^{k_{m}l}\xhc^{k_{m+1}}+\dots
\end{displaymath}
where the dots indicate further terms which contain less than $m\!+\!1$
$\xhc^{k}$-tensors. Equation (\ref{RL-condition}) then yields
$-\half(m\!+\!3)=0$, which is again a contradiction.

One can now see that as we subsequently replace two $\xhc^{k}$-terms by
one $\delta$-tensor, equation (\ref{RL-condition}) gives that
$-\half(m\!+\!7\!-\!2p)=0$, where $p$ is the total number of
$\xhc^{k}$-terms exchanged as compared to (\ref{RL-test}). But this
shows that $m$ has to be an odd integer in contradiction to our
demand that $m$ be an even number. As a result no type A tensor can have
the maximal number of momentum variables in the fermionic part of the
Runge-Lenz vector.

We refrain from repeating the argument for type B tensors since the
result is the same: The time-derivative in acting on these tensors
increases the number of $p^{k}$-variables by one and there is no way
to cancel all these extra unwanted terms against each other. We can
therefore safely conclude that even the introduction of higher orders
of momentum does not change the overall result, namely that a
supersymmetric version of the Runge-Lenz vector does not exist.

\section*{Acknowledgements}
R.H. gratefully acknowledges the support of EPSRC and the
Studienstiftung des Deutschen Volkes. The author would further like to
thank E. Gozzi for communications.

\end{document}